# First Programming Language: Visual or Textual?

**Mark Noone and Aidan Mooney [S, C]**
mark.noone@mu.ie, aidan.mooney@mu.ie
**Department of Computer Science**
**Maynooth University**
**Maynooth, Co. Kildare, Ireland**

| Please tick type of submission (see Call For Papers for more detail): | Tick (√) |
|---|---|
| **Research Paper (10 pages incl. literature review and methodology)** | √ |
| **Discussion Paper (10 pages incl. literature review)** | |
| **Extended Abstracts (2-3 pages in extended abstract format)** | |

S = Senior Author, C = Corresponding Author

## Abstract

In modern day society, the ability to code is a highly desirable skill. So much so that the current supply from third level institutes across the world does not meet the high demands of industry. One of the major issues is the low progression rates from first to second year in third level Computer Science courses with introductory programming courses proving to be a high contributing factor. This is something that needs to be addressed. One such way to address the issue is to get children involved and engaged with computing at young ages.

This paper describes a study undertaken that is the first step in a body of work that aims to garner the interest of potential Computer Science students at an early age. The study involves a comparison of two short courses; one based in Java and one based in Snap. The goal is to determine whether either of these languages is a better first programming language for students than the other, or if both are viable. These languages were chosen to allow for a comparison between a Visual Programming Language and a Textual Programming Language.

Feedback in the form of a survey will be used to gather the opinions of the students. This will provide data on issues such as which language was easier to learn and which language was preferred amongst others. Based on the outcomes of this study, a full-scale curriculum will be developed in the coming year. The outcomes of this study will help to establish which is the best programming language to suit the learning needs of students.

## Keywords

# 1. Introduction and Motivation

Computer Science as a third level subject has a history of volatility. While it has generally provided a very high graduate to employment ratio, it has often struggled with retaining students in the early years, particularly from first to second year (Quille, Bergin, and Mooney, 2015). According to a study undertaken by the Irish Times newspaper in 2016, "about one-third of (Irish) Computer Science students across all institutes of technology are dropping out after first year in college" (O'Brien, 2016). This is something that researchers and educators continually try to mitigate.

The motivation for this study relates closely to the results of a Systematic Literature Review undertaken by the authors (Noone and Mooney, 2017). This review pertains to the concept that tackling the task of introducing students to Computer Science, and more specifically programming, should be done at an early age in order to best pique and maintain their interest. In 2018, Computer Science will be introduced as an optional subject at Leaving Certificate level in Ireland (O'Brien, 2017). A short coding course is also an option for students taking the Junior Cycle (Curriculum Online, 2017), which was introduced in 2017.

This paper will examine the perception of both Visual Programming Languages (VPL's) and Textual Programming Languages (TPL's) amongst students aged between 10 and 18. Specifically, two short courses were developed; one in the programming language Snap (VPL) and one in the programming language Java (TPL). These courses were then delivered at a summer camp in the Computer Science department at Maynooth University. In order to determine the perception and outcomes, a survey was administered and analysed. The results from this survey will help to guide development of a long form curriculum to be delivered to secondary school students in the future. The overall goal is for this to be a "best bet approach" to keeping younger students interested in undertaking a Computer Science course at third level.

# 2. Background and Related Literature

The study presented in this paper follows on from a Systematic Literature Review that was undertaken between October 2016 and March 2017 (Noone and Mooney, 2017). This review aimed to determine the value in teaching VPL's as well as defining a "best choice" of VPL and TPL for teaching young learners Computer Science and programming. The results of this SLR heavily guided the study detailed in this paper, particularly in terms of language choices.

The languages that were decided upon to be used in this study were Snap and Java. Snap was chosen as an alternative to the popular VPL Scratch as Snap is a clone of Scratch that offers some more advanced options which will be useful for future work in the development of a curriculum. Java was chosen because, according to the TIOBE index (TIOBE, 2017), Java is the most commonly used programming language in the world. This includes an aggregate of both educational and industry based popularity.

A major finding of the literature review was that teaching methodologies are often more important than the actual languages of choice. While there is a multitude of literature that pointed to these languages being the "best choice", there were some key points that added further weight to using them as the languages of choice in this study. Given that language choice isn't the most important aspect of teaching, picking a language that is widespread has clear advantages for both teacher and learner. There is also more established material available for teaching Java and Snap. Snap provides all the functionality of Scratch along with the ability to add additional features.

To further this point, a survey conducted in the USA in 2011 (Davies, Polack-Wahl and Anewalt, 2011) investigated 371 institutions of educations who were asked what language they used for their beginner programming courses (CS1 or CS0). The results were that 48.2% of institutions had adopted Java. While this doesn't inform us of the current rates today, nor does it inform us of data outside of the USA, it still gives a strong idea of where the numbers have been in recent years.

In many countries, CoderDojo and other similar after school club organisations often use Scratch or another visual language as the backbone of their teaching. It is well established that VPL's are considered more fun for young learners to experiment with when compared to the more complex TPL's. The authors have had experience with running such a CoderDojo and have seen first-hand the effect learning Scratch, and other blocks based languages, has had on some students. Other researchers have examined this further by looking at retention of students (Armoni, Meerbaum-Salant and Ben-Ari, 2015). They found that after learning Scratch at an early age, the students that chose to progress to a Java/C# course in later years appeared to pick up information faster and grasp some of the tougher concepts earlier than their peers.

However, which of these two approaches makes the learning process easier for the student? This is the question we were hoping to answer. As mentioned earlier, language choice and approach is not as important as methodology; however, some elements of each language type might be easier for different age groups to learn. Due to this, a combination of both languages might be a good, if not the best, approach to take. A study undertaken in 2015 which involved teaching five weeks of Snap and five weeks of Java found that 58% of students thought Snap was easier to use (Weintrop and Wilensky, 2015). Some students reported that the blocks approach was easier to read. Some drawbacks were also noted; for example, blocks were identified as being less powerful giving less implicit customisation (but this isn't a huge issue for CS1).

These are but a few examples of why these languages were chosen. In general, no matter what language or tool is chosen it will have both positives and negatives. For this study, the two languages were chosen due to pedagogical evidence of their success.

## 3. Methodology

This section will describe the implementation of the two short courses; one in Java and one in Snap. This includes the full development of each course as well as a survey and the overall plan for implementing the study.

**3.1 Curriculum Development**

Once the two languages had been decided on, work could commence on developing a short course style curriculum in both. The goal was to create two courses that were close to identical in terms of content and level of difficulty, while still managing to showcase the important elements of each respective language. The courses would be designed to be delivered to students aged 10-18 in 90-minute sessions. As well as the author, multiple demonstrators would be employed at the summer camp to provide help to the students whenever they struggled. This would allow for even detailed topics to be covered in a very short timeframe. In terms of content, both courses would cover the topics presented in Table 3.1.

For Java, the course closely followed a shortened, but expedited version of the CS1 course delivered in the Computer Science department at Maynooth University. For Snap, elements were taken from the "Beauty and Joy of Computing course" (BJC, 2017) as well as from personal experiences with Scratch. The key element that allows these courses to be delivered in such a short timeframe (compared to usually spending weeks learning these topics) lies in the expectations. The students are not expected to become experts in the material. We simply aim to give them an overview of what programming entails and introduce some threshold concepts.

| *Language Tools (BlueJ / Snap)* | *Java boilerplate / Snap run blocks* | *"Hello World" for the language* | *Selection statements (if/else)* |
|---|---|---|---|
| *Basic Math* | *Variables* | *Loops* | *An advanced topic* |

**Table 3.1 – Course Topics**

At the end of each topic in a course, an exercise would be displayed on the slides giving the students a chance to trial what they have learned. The students' copy of the material would not contain the answers to these exercises. Once the students had sufficient time to work on the exercise, the answer would be given on the lecturer's copy of the slides on screen so they could see the optimal solution.

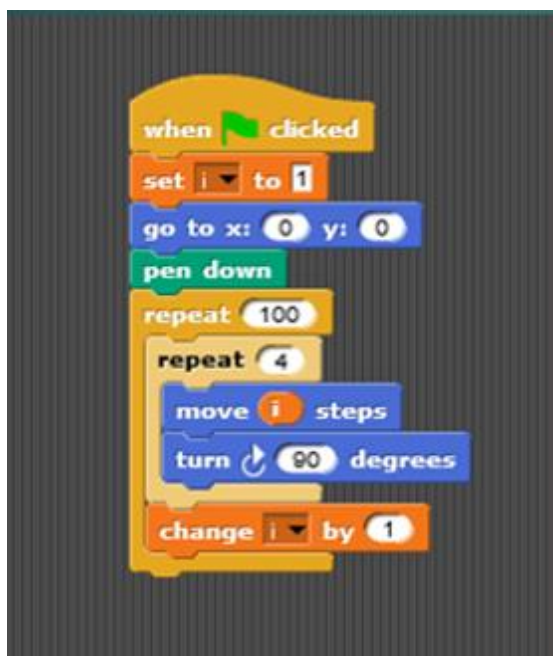 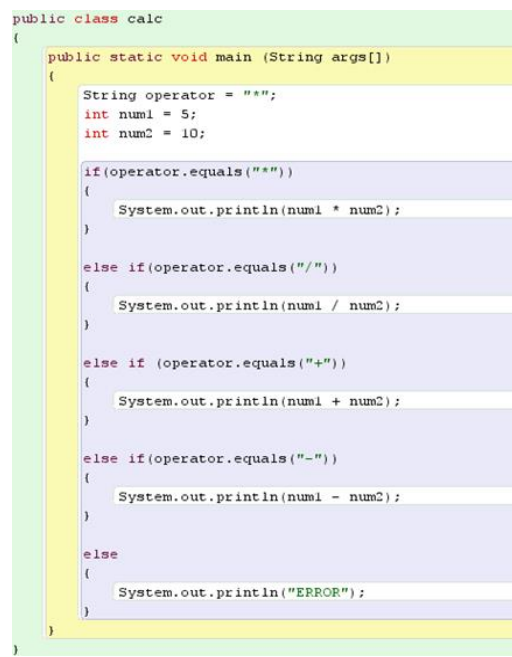

**Figure 3.1(a) – Snap Advanced Code**  **Figure 3.1(b) – Java Advanced Code**

For the final section of each course, the advanced topic was covered. This is the only element of the courses that would be significantly different across each course. For Snap, the concept of drawing shapes was chosen since it would utilise a part of everything they had learned so far and also demonstrate some nice animation features of the software (See Fig. 3.1(a)). For Java, the concept of creating a very basic calculator was chosen. Again, this would combine everything they had learned together, while showing something functional that they would understand (See Fig. 3.1(b)).

**3.2 Pilot Tests**

Before the commencement of the main testing phase at the Computer Science Summer Camp, a pilot test for each course was conducted. The Snap course was tested on a cohort of approximately 20 Junior Cycle students. These students were attending Maynooth University to experience taster courses in multiple disciplines. The material was well received based on anecdotal feedback. Many students followed along with the course material, while some others lost focus. This was not a major concern as it would be expected with a group who had not decided themselves to attend the course.

The Java course was also tested on a small group of Senior Cycle students by a member of our research group. The anecdotal feedback from them was generally positive. Some students struggled with the loops concept (a threshold concept within programming). This was rectified in the final course by moving it to later in the teaching process for those who wanted a challenge after completing the main phase.

The feedback received from both of these pilot tests was vital as it helped to verify that the timing of the courses was correct as well as revealing some enhancements that were needed in the material.

**3.3 The Survey**

Since there would be no official testing of the students, feedback relating to the success or failure of the courses would come from the results of a survey. This survey was compiled, using Google Forms, after the initial pilot tests were completed. The goal of the survey was to learn about how the students felt about the courses; how the courses compared was a key element of this. The questions that were decided upon for the survey can be seen in Appendix A.

These questions would help to gather feedback from the students with as little bias as possible. It would allow them to express their preferences within each individual course as well as make a fair comparison of the two courses. The ability to filter the results by age and gender is also key to determining if there are any preference patterns. In line with ethical requirements, consent forms were given to parents before the commencement of the camp to ensure data collection and possibly publication was allowed by their parents. On the day, consent from the students was also collected.

**3.4 Course Delivery During the Summer Camp**

In June 2017, the departmental summer camp began. The camp is broken down into three separate weeks where students can choose to do any of the weeks individually or all three weeks. All the content was unique in each week. The Java and Snap sessions were both scheduled to run on the same day in week two. In total, there were 35 students sitting the courses on the day with ages varying from 10 to 18 years of age.

The Snap session was the first session delivered. The students all chose a PC which had the Snap website preloaded as well as a copy of the material opened. After a short introduction, the slides were presented and delivered at a slow pace. The majority of students followed along with no issues, with minimal assistance from the demonstrators. Some students were on the wrong track with some of the exercises, but understood the answer once it was shown on the screen. This session ran for 90 minutes.

After a short 30-minute lunch break, the students returned and immediately started working with Java. Java was chosen to go second due to the perceived extra difficulty it would present. Since the material of the two courses mostly matched, they would only be learning the syntax of Java in the first part of the Java course rather than both the concepts and the syntax. To further assist with the learning of Java, some parts of the exercises were live coded after the students made their attempt rather than being static on the screen. After the completion of the Java course, the survey was immediately administered while all the material was still fresh in their minds. If there was more time available, another study would have been ran using Java first as well.

## 4. Results

This section will look at the analysis of the data gathered at the CS Summer Camp as well as anecdotal feedback from the students about their opinions of the courses.

## 4.1 Survey Analysis

All 35 students who were present on the day of the study provided a response to the survey. The demographic of the participants varied slightly with 88.6% of the participants being male and 11.4% being female. In terms of the age groups, 8.6% were between 10 and 12, 68.6% were between 13 and 15, and 22.9% were 16 or older. To examine if these demographics show patterns in their perceptions of the different languages and styles, most feedback will be broken down in relation to age groups.

The most encouraging outcome from the survey was that 88.6% of respondents said they wanted to learn more programming in the future. The other 11.4% said that they "maybe" want to learn more in the future. This shows that the perception of younger learners towards the topic is very positive.

The main question that arose from this study is whether one language was perceived to be harder than the other or not. The results of this question can be seen in Figure 4.1(a) and Figure 4.1(b). The mean difficulty rating for Snap was 3.57/10 while Java had a mean difficulty rating of 6.94/10. Based on a one-tail paired two sample t-test, this represents a clear statistical inference that Java was harder to learn than Snap ($p = 6.8E-10$). We wanted to check how this breaks down across the age groups? Is there a clear upward trend of the languages getting easier as you get older? From the results of this study, there seems not to be a clear trend, as can be seen in Figures 4.2(a) and 4.2(b). Unsurprisingly, for all age groups, Java remained more difficult than Snap (10-12 year olds: p=0.0471, 13-15 year olds: p=1.03E-06 and 16+ year olds: p=0.0013).

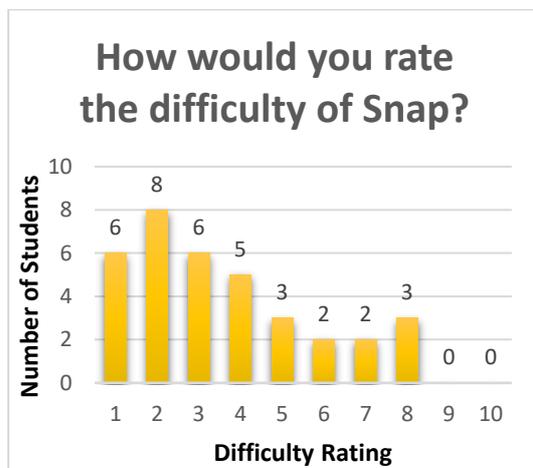 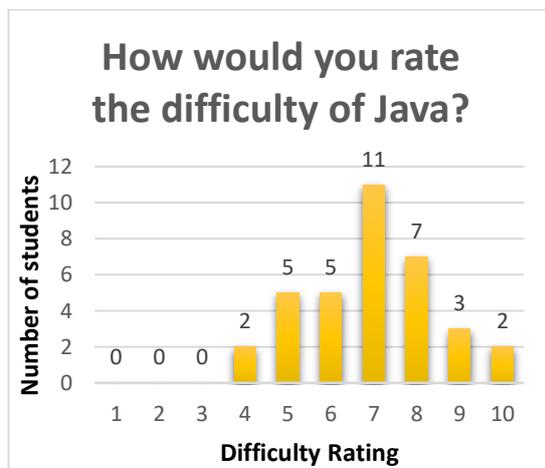

**Figure 4.1(a) – Snap Difficulty**     **Figure 4.1(b) – Java Difficulty**

This is not conclusive given that the number of students who fell outside the 13-15-year-old range was very low. Given that the courses were created with the same core content (with the exception of the one advanced topic in each), no external bias is being added to the difficulty ratings. This is purely based on the syntax and content of the languages themselves and how they compare to each other.

To look deeper, the students were asked what they considered to be the hardest aspect of each course. For Snap, 34.3% said variables were, 28.6% said drawing shapes and 22.9% said loops. For Java, 68.6% said making the calculator, 17.1% said loops and 11.4% said selection statements. Putting the shapes and the calculator on this section of the survey was an oversight as they involved using all the core elements of the course to make. However, this still gives us a good idea of what students were finding difficulty with and were mostly in the expected places.

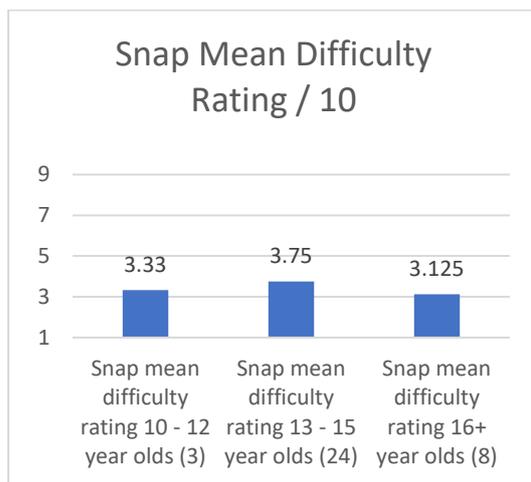
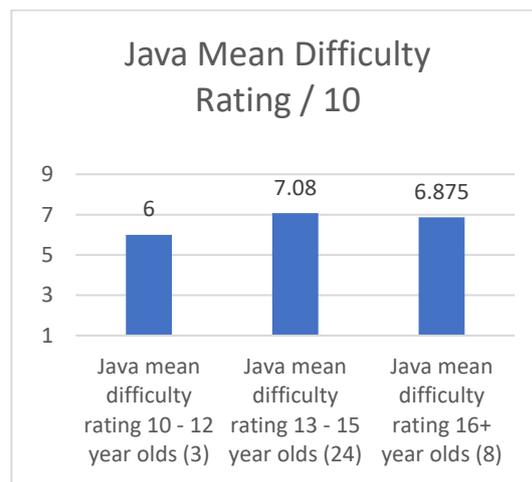

**Figure 4.2(a) – Snap Difficulty by Age**       **Figure 4.2(b) – Java Difficulty by Age**

Finally, when asked whether they preferred one of the styles of programming (text based or visual based), 37.1% of students said they preferred text based and 31.4% said visual based and 31.4% said that they had no preference. However, analysing the age brackets for this question it was observed that 13-15 year olds had a larger preference for the visual based language (41.6%) over the text based language (29.2%) with 19.2% having no preference. Conversely, 16+ year olds overwhelmingly preferred text based programming (62.5%) with 0% choosing the visual based option and 37.5% having no preference. This gives some credence to the theory that younger students enjoy visual programming more and older students' prefer textual programming.

## 4.2 Anecdotal Feedback

As well as the numerical data presented in Section 4.1, the students could make some optional comments on the course related to their favourite and least favourite elements as well as anything else they wished to add. When asked if they enjoyed each course, 60% of students responded that they did (for both courses). Similarly, 34.3% (Snap) and 25.7% (Java) said that they though the courses were "OK". The remaining 5.7% (Snap) and 14.3% (Java) did not enjoy the courses. In terms of a preferred course, 51.4% of the students preferred the Snap course with 48.6% preferring Java.

In terms of students' favourite and least favourite things in each course, some students chose not to respond, and others gave spoiled answers. For favourite things, there were 30 valid answers while for least favourite things there were 24 valid answers. These were then summarised into categories and the results can be seen in Figure 4.3(a) and Figure 4.3(b).

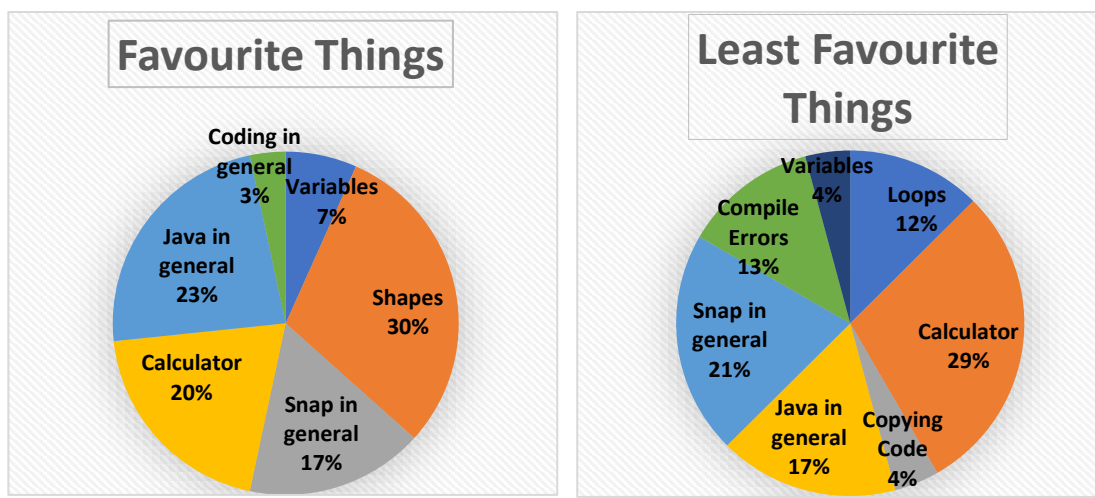

Figure 4.3(a) – Students responses to their favourite things in the courses

Figure 4.3(b) – Students responses to their least favourite things in the courses

A few other key anecdotal comments from the survey included:

- *"Make the programming bit more interesting"* – This was a fair comment. It is difficult to find the balance between learning and fun in an introductory course.
- *"Spend more time on Java"* – Given the higher difficulty of Java, it is completely fair that students would need more time to learn the Java material.
- *"Overall it was really interesting and enjoyable!"*

## 5. Conclusions and Future Work

The results of this study, even though the number in attendance was reasonably small and a lot of the feedback was anecdotal, are still important. When and why one might use a VPL as a teaching tool are questions that are often asked and rarely answered. This study has shown that a language like Snap has as much potential for learning as Java, provided the target audience is correct.

Some elements of the study could have been done differently. The time available for delivery of the course was not ideal with some students feeling like they needed longer for certain concepts. There were some minor mistakes in the survey such as asking the students what the most difficult element was and including the element that encompassed everything. These issues will be ironed out for future studies.

Without a doubt, both languages deserve further investigation. In the coming months, these curricula will be scaled up into a multi-week course for delivery in secondary schools. The target audience will be Transition Year students (approximately 15 years old). The longer versions of these courses will spread out the material more, allowing for time to absorb the content fully. New and more advanced elements will also be added into the content of course. Assessment will involve both surveys and quizzes / tests. Finally, the courses will be delivered to multiple groups with some learning Java first and others learning Snap first. This will provide a clearer picture.

This new course will align perfectly with the introduction of Computer Science as a Leaving Certificate subject in 2018. The timing could not be better for testing out what does and does not work at an entry level. It is very important to be aware that the choice of language is not always the most important choice but how it is delivered to the students. With that said computer programming is an important skill and knowing which type of computer language engages students at different ages can help improve interest in the subject. It is hoped that with the aid of these courses, more students can be guided towards engaging with Computer Science during their school life and eventually leading them to college courses and careers in the field.

## 6. Acknowledgments

This work was completed with the help of funding from the John and Pat Hume scholarship, Maynooth University.

# References

Armoni, M., Meerbaum-Salant, O., & Ben-Ari, M. (2015). From scratch to "real" programming. *ACM Transactions on Computing Education (TOCE)*, *14*(4), 25.

BJC (2017). Beauty and Joy of Computing an AP CS Principles course. Retrieved August 12, 2017, from bjc.edu.org

Curriculum Online (2017, June). Short Course Coding Specification for Junior Cycle. Retrieved August 12, 2017, from https://curriculumonline.ie/getmedia/cc254b82-1114-496e-bc4a-11f5b14a557f/NCCA-JC-Short-Course-Coding.pdf

Davies, S., Polack-Wahl, J. A., & Anewalt, K. (2011, March). A snapshot of current practices in teaching the introductory programming sequence. In *Proceedings of the 42nd ACM technical symposium on Computer science education* (pp. 625-630). ACM.

Noone, M., & Mooney, A. (2017, October). Visual and Textual Programming Languages: A Systematic Review of the Literature. ArXiv e-prints Journal. Retrieved October 5, 2017, from https://arxiv.org/abs/1710.01547

O'Brien, C. (2017, February). Computer science to be fast-tracked onto Leaving Cert. Retrieved August 12, 2017, from https://www.irishtimes.com/news/education/computer-science-to-be-fast-tracked-onto-leaving-cert-1.2964672

O'Brien, C. (2016, January). Concern over drop-out rates in computer science courses. Retrieved August 12, 2017, from https://www.irishtimes.com/news/education/concern-over-drop-out-rates-in-computer-science-courses-1.2491751

Quille, K, Bergin, S., & Mooney, A. (2015). PreSS#, A Web-Based Educational System to Predict Programming Performance. International Journal of Computer Science and Software Engineering (IJCSSE), 4 (7). pp. 178-189. ISSN 2409-4285.

TIOBE (2017, August). TIOBE Index for August 2017. Retrieved August 13, 2017, from http://www.tiobe.com/tiobe-index/

Weintrop, D., & Wilensky, U. (2015, June). To block or not to block, that is the question: students' perceptions of blocks-based programming. In *Proceedings of the 14th International Conference on Interaction Design and Children* (pp. 199-208). ACM.

# Appendix A - Survey

1. What age are you?
   - 9 or younger
   - 10 – 12
   - 13 – 15
   - 16 or Older

2. What gender are you? (Male, Female)
3. How would you rate the difficulty of Snap? (1-10)
4. How would you rate the difficulty of Java? (1-10)
5. What was the hardest aspect of Snap to learn?
   - Variables
   - Selection (if / else)
   - Loops
   - Drawing Shapes

6. What was the hardest aspect of Java to learn?
   - Variables
   - Selection (if / else)
   - Loops
   - Making a Calculator

7. Did you enjoy the Snap course? (Yes / No / It was OK)
8. Did you enjoy the Java course? (Yes / No / It was OK)
9. Which course did you prefer? (Snap / Java)
10. What was your favourite thing from either course?
11. What was your least favourite thing from either course?
12. Which style of programming do you prefer?
    - Text (Java)
    - Blocks (Snap)
    - Both are equally good!

13. Would you like to learn more programming in the future? (Yes / No / Maybe)
14. Any other comments?